%% file: arxiv.tex
\documentclass[conference]{IEEEtran}
\IEEEoverridecommandlockouts
\usepackage{cite}
\usepackage{amsmath,amssymb,amsthm}
\usepackage{bbm}
\usepackage{algpseudocode}
\usepackage{graphicx}
\usepackage{textcomp}
\usepackage{mathtools}
\usepackage{xcolor}
\usepackage{algorithm}
\usepackage{bm}

\def\BibTeX{{\rm B\kern-.05em{\sc i\kern-.025em b}\kern-.08em
    T\kern-.1667em\lower.7ex\hbox{E}\kern-.125emX}}
\usepackage{url}

\usepackage{tikz}
    \usetikzlibrary{shapes.arrows}
    
\usepackage{subfig}
\usepackage{pgfplots}
\pgfplotsset{
compat=1.3,
legend style={font=\footnotesize, fill opacity=0.7,  draw opacity=1, text opacity=1, draw=white!15!black, legend cell align=left, align=left}, 
width=6cm, 
height=6cm,
yminorticks=false,
xminorticks=false,
title style={font=\small},
tick style={color=black},
tick label style={font=\small},
grid style={line width=.1pt, draw=gray!20},
major grid style={line width=.1pt,draw=gray!20},
}
\usepgfplotslibrary{colorbrewer}
\usepgfplotslibrary{groupplots}
\usepgfplotslibrary{fillbetween}
\usetikzlibrary{decorations.pathreplacing}
\usetikzlibrary{plotmarks}
\usepackage[]{fancyhdr}
\usetikzlibrary{patterns}
\pgfplotsset{every tick label/.append style={font=\footnotesize}}

\usepackage{tikz}
\usetikzlibrary{shapes, arrows, positioning}

\usepackage{glossaries}
\input{acronyms.tex}

\algtext*{EndWhile}
\algtext*{EndIf}
\algtext*{EndFor}

\definecolor{color0}{HTML}{00429D}
\definecolor{color1}{HTML}{844D99}
\definecolor{color2}{HTML}{C3608E}
\definecolor{color3}{HTML}{EF8078}
\definecolor{color4}{HTML}{FFB047}
\definecolor{darkslategray38}{RGB}{38,38,38}
\definecolor{darkblue}{HTML}{00429D}
\definecolor{darkgreen}{HTML}{005c00}
\definecolor{gold}{HTML}{D4AF37}
\definecolor{darkred}{HTML}{910000}

\makeatletter

\title{Transition Dynamics Matter: Hidden Markov Scoring of Epileptogenic Channels in SEEG
\thanks{Part of this work was supported by the European Union under the Marie Skłodowska-Curie Actions (MSCA) Postdoctoral Fellowships programme of Horizon Europe, Grant Agreement No. 101270737 (HYPHEN).}}

\author{\IEEEauthorblockN{Giulia Tacconi$^{\dagger}$, Federico Mason$^{*}$, Lorenzo Ferri$^{\ddagger}$, Elena Pasini$^{\ddagger}$, Lidia Di Vito$^{\ddagger}$, Matteo Martinoni$^{\ddagger,\S}$,\\
Lara Alvisi$^{\ddagger,\S}$, Anna Scarabello$^{\S}$, Luca Zanuttini$^{\S}$, Francesca Bisulli$^{\ddagger,\S}$, and Andrea Zanella$^{\dagger}$}
\IEEEauthorblockA{$^{\dagger}$Department of Information Engineering, University of Padova, Via G. Gradenigo 6/B, 35131, Padua, Italy\\
$^{*}$Institut de Neurosciences des Systèmes, Aix-Marseille University, Boulevard Jean Moulin 27, 13005, Marseille, France\\
$^{\ddagger}$IRCCS Institute of Neurological Sciences of Bologna, Via Altura 3, 40139, Bologna, Italy \\
$^{\S}$Department of Neuromotor and Biomedical Sciences, University of Bologna, Via Massarenti 9, 40138, Bologna, Italy \\
$^{*}$Corresponding author's email: federico.mason@univ-amu.fr}
}

\begin{document}

\maketitle

\begin{abstract}
Stereo-electroencephalography (SEEG) represents the most effective technique for the detection of the Epileptogenic Zone (EZ) in drug-resistant epilepsy.
This procedure produces a multi-dimensional signal with hundreds of components, named \emph{channels}, each associated with a different brain location. 
In this work, we propose a joint Hidden Markov Model (HMM), in which each SEEG channel is described by a latent Markov chain, with states shared across the network and channel-specific transition probabilities.
Accordingly, we develop a detection algorithm that identifies the channels belonging to the EZ by quantifying, through the Kullback-Leibler divergence, the difference between their transition probabilities in the inter-ictal period and at seizure onset.
We evaluate our framework in SEEG recordings from $30$ patients with drug-resistant epilepsy and compare it with a benchmark that operates in the same observation space but does not consider a latent-state structure.
Compared against medically validated EZ definitions, our study shows that modeling channel-specific temporal transitions can improve EZ localization and offer a data-driven framework to investigate the latent dynamics underlying seizures.
\end{abstract}

\begin{IEEEkeywords}
Stereo-electroencephalography, drug-resistant epilepsy, hidden Markov model, abnormality detection 
\end{IEEEkeywords}

\glsresetall

\section{Introduction}
\label{sec:intro}

\begin{tikzpicture}[remember picture, overlay]
\node[draw, minimum width=4in, font=\small, text=black] at ([yshift=-1cm]current page.north)  {This paper has been submitted to the IEEE International Conference on Communications (ICC) 2027.};
\end{tikzpicture}

In drug-resistant epilepsy, surgery is the treatment with the highest likelihood of achieving long-term seizure freedom \cite{rosenow2001}.
The goal is to resect the \gls{ez}, defined as the cortical area indispensable for the generation of seizures.
Accurate localization of the \gls{ez} is then essential to maximize the probability of seizure freedom after surgical intervention while preserving the eloquent cortex.
When noninvasive evaluation is not conclusive, clinical practice involves the use of invasive techniques such as \gls{seeg} to identify the \gls{ez}.
This procedure is based on implantation of needle electrodes, which record electrical activity from the deep brain structures with high temporal resolution, although with a limited spatial sampling \cite{isnard2018}.

The outcome of \gls{seeg} monitoring is a multi-dimensional signal with hundreds of components, named \emph{channels}, each associated with a different recording site.
These signals provide fundamental information about both the \emph{ictal} periods, i.e., seizures, and \emph{inter-ictal} periods, i.e., the intervals occurring between consecutive seizures.
In a clinical context, the analysis of \gls{seeg} signals is primarily based on visual interpretation, focusing on the detection of the channels that change most prominently during the transition to the ictal phase \cite{frauscher2026}.
Such a procedure requires the participation of highly specialized neurologists and can still be subject to interpretative errors due to the complexity of the signals and the necessity of integrating spatial and temporal information.

In recent years, \gls{seeg} interpretation has been supported by quantitative techniques with the aim of providing new biomarkers of epileptogenic activities.
In particular, the most common biomarker of seizure onset is given by \gls{lvfo}, i.e., low-amplitude activity spanning \emph{beta} and \emph{gamma} frequencies ($f\geq14$ Hz) \cite{wang2022diagnostic}.
To detect \gls{lvfo}, it is common to map \gls{seeg} signals to their time-frequency representations via the short-time Fourier transform, and then quantify the energy increase in high-frequency bands.
This is the approach used by the \gls{ei} \cite{bartolomei2008}, which is the most well-established technique to assist clinicians during \gls{seeg} review.
On a practical level, the \gls{ei} characterizes each \gls{seeg} channel through \emph{local} features, i.e., quantities computed from the target channel at a specific time point, regardless of the rest of the \gls{seeg} network.
However, seizure generation is recognized as a network phenomenon, and looking at channels as independent processes may be insufficient for the \gls{ez} delineation \cite{bartolomei2017defining}.

More advanced models treat \gls{seeg} signals as graphs, where nodes denote recording sites, and edges denote inter-channel connectivity.
Recent studies report that seizure onset is characterized by a reduction in connectivity for epileptogenic channels \cite{mason2026desynchronization}, followed by a hyper-synchronization during the propagation of epileptic discharges \cite{cheng2025localization}.
This pattern is well explained by the \emph{inter-ictal suppression} hypothesis, which proposes that the \gls{ez} is restrained by surrounding structures during the inter-ictal period \cite{doss2024interictal}.
However, approaches in the literature often consider \emph{time-resolved} graphs, i.e., sequences of graphs computed on successive time windows, and analyze each graph independently, without modeling the fact that \gls{seeg} signals span both \emph{space} and \emph{time}.
As suggested by the inter-ictal suppression hypothesis, seizures are not abrupt events but result from the brain evolving from a pre-ictal condition, where the \gls{ez} is inhibited, to an ictal condition, where the \gls{ez} takes control over the brain \cite{kayabas2024transition}.

A suitable approach for modeling the evolution of brain states is given by \glspl{hmm}, a probabilistic framework in which a sequence of observations is generated by latent states evolving as a Markov chain \cite{rabiner1989tutorial}.
This method separates the emission process (observations given states) from the transition process (temporal evolution among states).
In the literature, \glspl{hmm} have been used to differentiate between inter-ictal and ictal periods, enabling the development of seizure prediction methods \cite{baldassano2016novel} and the mapping between non-invasive features and intracranial patterns \cite{zhang2021dynamic}.
However, these previous works focused on modeling the whole-brain evolution and did not consider the possibility of modeling the latent states that characterize each brain location.

In this work, we develop a new framework to characterize \gls{seeg} signals,
explicitly modeling the transition dynamics that characterizes single channels.
We hypothesize that the generation of epileptic discharges can be represented through a transition between latent states, leading the channel from a baseline to abnormal (ictal-like) dynamics.
Therefore, the ease with which a channel transitions towards ictal-like states differentiates epileptogenic structures from non-epileptogenic ones.
Practically, we assume that each channel evolves according to a latent Markov chain, with each state associated with an observation pattern comprising local, e.g., spectral energy, and spatial features, e.g., functional connectivity.
The state occupied by a channel at a given time point determines not only the distribution of its features, but also the probabilities of transiting towards other states.

Our model assumes that the number of states and their emission probabilities are shared throughout the \gls{seeg} signal, while each channel is characterized by its own transition probability matrix.
In addition, the transition probabilities are not static but change when the brain changes its macro-conditions and, e.g., enters the ictal phase.
Since states and emissions are shared, the transition matrix is the only channel- and condition-specific parameter, and we can compare different channels, or different conditions of the same channel, by computing the \gls{kl} divergence between the corresponding transition probability matrices.
Hence, we assume that epileptogenic channels can be detected by looking at those channels that more prominently change the transition probabilities at the onset of the seizure. 

The main contributions of this study are the following:
\begin{itemize}
\item We propose a joint \gls{hmm} framework to characterize \gls{seeg} signals, where each channel is associated with its own transition matrix, while all channels share the same latent states and emission distributions;
\item We develop a detection algorithm that identifies the channels belonging to the \gls{ez} by computing the \gls{kl} divergence between the transition matrices estimated  during the inter-ictal period and at the beginning of the seizure;
\item We evaluate our methodology on a dataset of $30$ patients with drug-resistant epilepsy who underwent \gls{seeg} monitoring at the IRCCS Institute of Neurological Sciences of Bologna, using medically validated labels as ground truth for the \gls{ez} definition;
\item We compare the performance of our method against a benchmark approach, which characterizes each channel only through static emission probabilities, without considering the channel's temporal evolution.
\end{itemize}

Our results indicate that explicitly modeling the transition dynamics at the level of individual \gls{seeg} channels, as done by our framework, can improve the localization of the \gls{ez} with respect to an approach based on time-resolved features.
This may provide new solutions to investigate the latent dynamics behind seizure generation in drug-resistant epilepsy.

\section{Signal Model}
\label{sec:model}

We model an \gls{seeg} signal as a multivariate discrete-time series $\mathbf{s}[n]$, $n = 0, \dots, N-1$, with $s_i[n] \in \mathbb{R}$ denoting the value of channel $i \in \mathcal{C}$ at sample $n$.
Given the sampling frequency $f_s$ and the duration of the signal $T$, the total number of samples is $N=f_s \cdot T$.
We segment the signal into overlapping windows of duration $\tau$, with $\Delta \tau < \tau$ denoting the time shift between consecutive windows.

We choose $T$, $\tau$ and $\Delta\tau$ to be multiples of the sampling period $1/f_s$, so that each window contains an integer number $M = f_s \cdot \tau$ of samples, and consecutive windows are spaced apart by $K=f_s \cdot \Delta \tau$  samples. Therefore,  the total number of windows is $W = \lfloor \frac{T - \tau}{\Delta \tau} \rfloor + 1$, and each window $w\in\{0,1,...,W-1\}$ contains the samples from $w \cdot K$ to $w \cdot K+M-1$.
We denote by $x_i[w,m]$ the $m-th$ sample of the channel $i$ within the $w$-th window, i.e., $x_i[w,m]=s_i[wK+m]$, with $m\in\{0,1,...,M-1\}$.

We consider that the signal comprises both a baseline (inter-ictal) and seizure (ictal) condition, and denote by $w^*$ the window associated with the onset of the seizure. 
Each channel $i \in \mathcal{C}$ and window $w \in \left[0, W - 1\right]$ is associated with a vector of features $\mathbf{o}_i[w] \in \mathbb{R}^F$, characterizing the \gls{seeg} signal for that specific channel and window.
As formalized in the following section, the vectors $\left\{\mathbf{o}_i[w]\right\}$ constitute the observations of our joint \gls{hmm} model and take values in the same space for all channels $i \in \mathcal{C}$.


We consider a set $\mathcal{B}$ of frequency bands, with $|\mathcal{B}| = B$.
Each band $b \in \mathcal{B}$ is defined by the frequency interval $[f^b_{\text{low}}, f^b_{\text{high}})$, expressed in Hz, with $0 \le f^b_{\text{low}} < f^b_{\text{high}} \le f_s/2$.
The \gls{dft} of $x_i[w, m]$ is
\begin{equation}
X_i[w, k] = \sum_{m=0}^{M-1} h[m] \cdot x_i[w, m] \cdot e^{-j 2\pi k m / M},
\end{equation}
where $k = 0, \dots, M-1$ corresponds to the frequency $k \cdot f_s / M$, and $h[\cdot]$ is a Hanning window of length $M$.
Hence, the energy of $x_i[w,m]$ in band $b \in \mathcal{B}$ is then defined as
\begin{equation}
E_i^b[w] = \frac{1}{M} \sum_{k=k^b_{\text{low}}}^{k^b_{\text{high}}} \left| X_i[w, k] \right|^2,
\end{equation}
where $k^b_{\text{low}} = \lceil f^b_{\text{low}} \cdot M / f_s \rceil$ and $k^b_{\text{high}} = \lceil f^b_{\text{high}} \cdot M / f_s \rceil - 1$ are the indices of the first and last frequency bins of $b$.

We band-pass filter $x_i[w, m]$ over each frequency band $b \in \mathcal{B}$, and denote by $x_i^b[w, m]$, $m = 0, \dots, M-1$, the samples of $x_i[w]$ after being filtered in $[f_{\text{low}}^b, f_{\text{high}}^b)$.
Then, we compute the instantaneous phase of $x_i^b[w, m]$ as the argument of its analytic representation:
\begin{equation}
\theta_i^b[w,m] = \arg\left( x_i^b[w, m] + j , \mathcal{H}{x_i^b[w, \cdot]}[m] \right),
\end{equation}
where $\mathcal{H}\{\cdot\}$ denotes the discrete Hilbert transform.
We then bin the phase values of the signal using the Sturges rule, so that the phase domain $[-\pi, \pi)$ is segmented into $L = \lceil \log_2 M \rceil + 1$ uniform bins.
The entropy of $x_i^b[w,m]$ is then given by
\begin{equation}
H_i^b[w] = - \sum_{\ell=0}^{L-1} P\left(\theta_i^b[w] \in \vartheta_{\ell} \right) \text{log}_2 \left( P\left(\theta_i^b[w] \in \vartheta_{\ell} \right) \right),
\end{equation}
where we denoted by $P\left(\theta_i^b[w] \in \vartheta_{\ell} \right)$ the \emph{empirical probability} that the samples $\theta_i^b[w,m]$ fall into bin $\vartheta_{\ell}$.

Finally, we exploit the instantaneous phase to estimate the connectivity between any pair $(i,j)$ of channels for each window.
Given a frequency band $b \in \mathcal{B}$ and the phase sequences $\theta_i^b[w, m]$, $\theta_j^b[w, m]$, $m = 0, \dots, M-1$, we define the \gls{mi} between $i$ and $j$ at window $w$ as
\begin{equation}
MI_{i,j}^b[w] = H_i^b[w] + H_j^b[w] - H_{i,j}^b[w],
\end{equation}
where $H_i^b[w]$ and $H_j^b[w]$ are the entropies defined above, and $H_{i,j}^b[w]$ is the joint entropy of the two phase sequences.
To project mutual information values onto single channels for each window $w$, we then compute the connectivity strength:
\begin{equation}
S_i^b[w] = \sum_{j \in \mathcal{C} \setminus {i}} MI_{i,j}^b[w].
\end{equation}
This quantity denotes how strongly a specific channel is interconnected with the rest of the \gls{seeg} network at the considered window.
The observation vector $\mathbf{o}_i[w]$ of channel $i$ in window $w$ is obtained by concatenating the energy, the entropy, and the strength over all frequency bands:
\begin{equation}
\mathbf{o}_i[w] = \left[ E_i^b[w], \cdots , H_i^b[w], \cdots , S_i^b[w] \right] \in \mathbb{R}^{3 B}.
\end{equation}
Notably, the energy and the entropy capture the \emph{local} properties of the channel, whereas the connectivity strength captures its \emph{spatial} properties.
We highlight that the model can be easily extended to other, more advanced features, as the observation space is not a constraint of our methodology.

\section{Joint Hidden Markov Model}
\label{sec:jointhmm}

\begin{figure}[!t]
  \centering
\includegraphics[width=0.9\columnwidth]{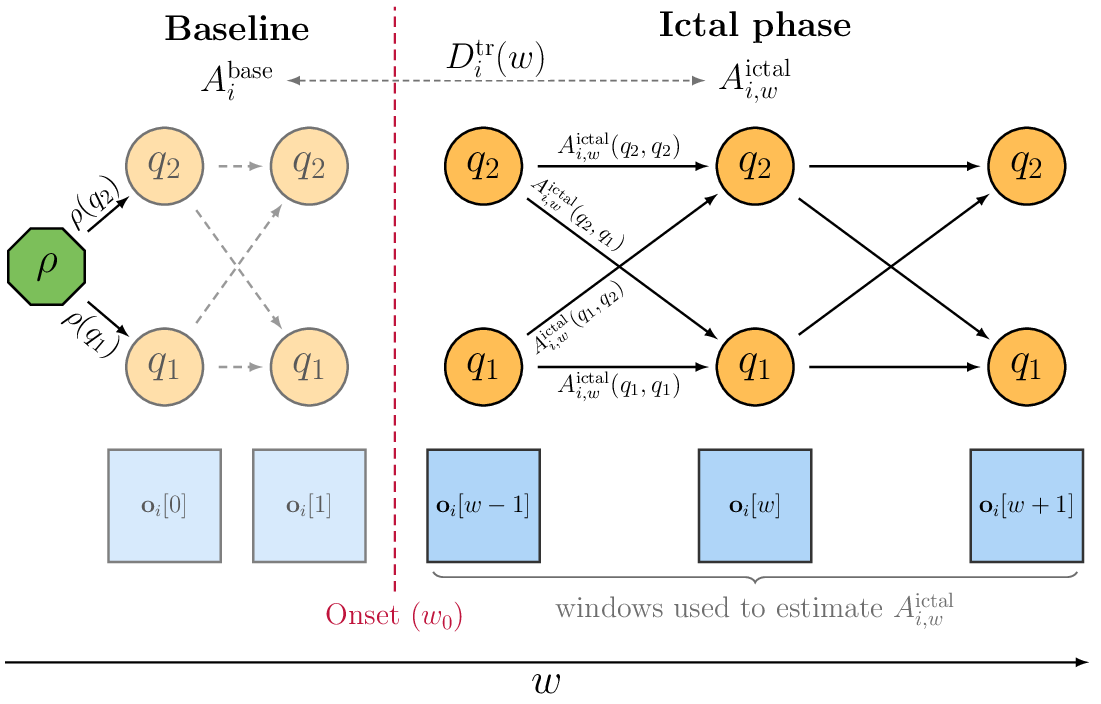}
  \caption{HMM associated with channel $i \in \mathcal{C}$, with $\mathcal{Z}=\{q_1,q_2\}$. \vspace{-0.0cm}}
  \label{fig:trellis}
\end{figure}

Given a \gls{seeg} signal $\mathbf{s}[n]$, we assume that the feature sequence $\mathbf{o}_i[0], \dots, \mathbf{o}_i[W-1]$ associated with each channel $i \in \mathcal{C}$ represents the observable outcome of a latent process that can be modeled via an \gls{hmm}.
Accordingly, $\mathbf{s}[n]$ can be modeled as $|\mathcal{C}|$ parallel \glspl{hmm}, each generating a sequence of latent states and associated observations.
A sketch of the \gls{hmm} associated with a channel is reported in Fig. \ref{fig:trellis}.

\subsection{Model Parameter Definition}

We denote by $z_i[w] \in \mathcal{Z}$ the latent state of channel $i$ in window $w$, where $\mathcal{Z}$ is a finite state space with $|\mathcal{Z}| = Z$ states. 
Hence, we assume that $z_i[w]$ is a first-order Markov chain, with stationary transition probabilities, and that the observation $\mathbf{o}_i[w]$ depends only on the current state $z_i[w]$.
Formally, the \gls{hmm} associated with channel $i \in \mathcal{C}$ is defined by the tuple $\left(A_i, \varphi, \rho \right)$, where:
\begin{itemize}
\item $A_i(z,q)=P(z_i[w+1]=q \mid z_i[w]=z)$ is the \emph{transition probability} that channel $i \in \mathcal{C}$ moves from state $z \in \mathcal{Z}$ to state $q \in \mathcal{Z}$, with $\sum_{q \in \mathcal{Z}} A_i(z,q) = 1$, $\forall$ $z \in \mathcal{Z}$;
\item $\varphi(\mathbf{o} \mid z)$, referred to as \emph{emission density}, reflects the probability of observing $\mathbf{o}_i[w] \in \mathbb{R}^{F}$ when channel $i \in \mathcal{C}$ is in state $z \in \mathcal{Z}$, and is identical for all channels;
\item $\rho(z)=P(z_i[0]=z)$ is the \emph{initial probability} that channel $i \in \mathcal{C}$ starts the process in state $z \in \mathcal{Z}$, with $\sum_{z \in \mathcal{Z}} \rho(z) = 1$, and is also identical for all channels.
\end{itemize} 

We note that, since the observations are continuous-valued, their statistical properties are expressed through probability density functions.
In this work, we assume that, for each state $z \in \mathcal{Z}$, the emission density $\varphi(\mathbf{o} \mid z)$ is a multivariate Gaussian with mean $\boldsymbol{\mu}_z \in \mathbb{R}^F$ and covariance $\Sigma_z \in \mathbb{R}_+^{F \times F}$, i.e., $\varphi(\mathbf{o} \mid z) = \mathcal{N}(\mathbf{o}; \boldsymbol{\mu}_z, \Sigma_z)$.
In particular, we consider $\Sigma_z$ to be a diagonal matrix\footnote{This assumption is common in \gls{hmm} frameworks, but it may be unrealistic in our setting, since the features associated with the same band are likely correlated.
It will be relaxed in future extensions of this work.}, which means that the features are conditionally independent given the current state $z \in \mathcal{Z}$.
Once all the parameters of the model have been defined, any sequence of observations $\mathbf{o}_i[0], \dots, \mathbf{o}_i[W-1]$ can be mapped to the most probable sequence of latent states $z_i[0], \dots, z_i[W-1]$ via the \emph{Viterbi} algorithm.

\subsection{Model Parameter Estimation}

Given a \gls{seeg} signal $\mathbf{x}$, we can estimate the parameters of the joint \gls{hmm} by means of the \gls{bw} algorithm, a special case of the \gls{em} algorithm \cite{rabiner1989tutorial}.
This approach iteratively seeks a maximum likelihood estimate of the model parameters, given the observed sequences.
To this end, it alternates an \emph{expectation step}, computing the posterior probabilities of the latent states given the observations and the model parameters, and a \emph{maximization step}, updating the parameters of the \gls{hmm} model based on these posterior probabilities.

In this work, we constrain the original \gls{bw} algorithm so that the emission densities and the initial probabilities are shared across channels.
For each signal $\mathbf{s}[n]$, we consider two stages: first, we consider the observation sequences of all channels as independent sequences sharing the same parameters, and estimate $\varphi(\mathbf{o} \mid z)$ and $\rho(z)$ via the original \gls{bw} algorithm.
Then, for each channel $i \in \mathcal{C}$, we run the algorithm to estimate the channel-specific transition matrix $A_i(z,q)$ by using only the observations of channel $i$ and keeping $\varphi(\mathbf{o} \mid z)$ and $\rho(z)$ fixed.
In particular, the second stage can be restricted to a subset of windows, so that $A_i(z,q)$ is estimated from the portion of the data of interest only.
This yields piecewise time-homogeneous transition probabilities, i.e., constant within a portion of the signal, for each channel.

For each channel $i \in \mathcal{C}$, we first estimate a baseline transition probability matrix $A_i^{\text{base}}(z,q)$ using all the windows of the inter-ictal condition, i.e., $w = 0, \dots, w^* - 1$.
We then estimate multiple ictal transition probability matrices $\left\{A_{i,w}^{\text{ictal}}(z,q)\right\}_{w}$, one for each window $w \geq w^*$ of the ictal phase.
We derive $A_{i,w}^{\text{ictal}}(z,q)$ using two different approaches: the \emph{cumulative} approach considers the observations from the onset of the ictal phase, that is, window $w^*$, up to the target time window $w$;
the \emph{sliding} approach consider the observations from the range $\left\{ w - L, \dots, w + L \right\}$, where $L$ is a hyperparameter determining the estimation horizon.
Finally, we denote by $\boldsymbol{\pi}_i^{\text{base}}(z)$ and $\boldsymbol{\pi}_{i,w}^{\text{ictal}}(z)$ the steady-state probability vectors of $A_i^{\text{base}}(z,q)$ and $A_{i,w}^{\text{ictal}}(z,q)$, respectively, i.e., the vectors $\boldsymbol{\pi}(z)$ that satisfy $\boldsymbol{\pi}(z) A (z,q) = \boldsymbol{\pi}(z)$, $\boldsymbol{\pi}(z) \mathbf{1} = 1$, and $\pi(z) \geq 0$ for $A(z,q) = A_i^{\text{base}}(z,q)$ and $A(z,q) = A_{i,w}^{\text{ictal}}(z,q)$.
These vectors represent the fraction of time that channel $i$ would spend in each state in the long run.

\subsection{Epileptogenic Channel Detection}
\label{subsec:scoring}

To quantify how much the dynamics of channel $i \in \mathcal{C}$ change at the onset of the seizure, we consider two different implementations of the \gls{kl} divergence.
In the first, referred to as \emph{transition KL}, corresponds to the \gls{kl} divergence between the rows of the transition matrices $A_i^{\text{base}}$ and $A_{i,w}^{\text{ictal}}$, weighted by the stationary distribution $\boldsymbol{\pi}_i^{\text{base}}$, obtaining
\begin{equation}
D_i^{\text{tr}}(w) = \sum_{z \in \mathcal{Z}} \boldsymbol{\pi}_{i}^{\text{base}}(z) \sum_{q \in \mathcal{Z}} A_{i,w}^{\text{ictal}}(z,q) \log \frac{A_{i,w}^{\text{ictal}}(z,q)}{A_i^{\text{base}}(z,q)}.
\label{eq:transition_kl}
\end{equation}

The second implementation, referred to as \emph{occupancy KL}, corresponds to the \gls{kl} divergence between $\boldsymbol{\pi}_i^{\text{base}}$ and $\boldsymbol{\pi}_{i,w}^{\text{ictal}}$:
\begin{equation}
D_i^{\text{occ}}(w) = \sum_{z \in \mathcal{Z}} \boldsymbol{\pi}_{i,w}^{\text{ictal}}(z) \log \frac{\boldsymbol{\pi}_{i,w}^{\text{ictal}}(z)}{\boldsymbol{\pi}_i^{\text{base}}(z)}.
\label{eq:occupancy_kl}
\end{equation}
By combining~\eqref{eq:transition_kl} and~\eqref{eq:occupancy_kl} with the cumulative and sliding strategies to estimate $A_{i,w}^{\text{ictal}}$, we obtain four variants of our framework.
To obtain an aggregate score for each channel, we sum the values of the \gls{kl} divergence over all the windows in the ictal phase, obtaining $D_i = \sum_{w=w^*}^{w_1} D_i(w)$.
This value denotes the deviation of channel $i$ from its baseline, representing a biomarker of channel epileptogenicity: the higher the score, the more likely the channel belongs to the \gls{ez}.

Finally, we define a \emph{benchmark} that characterizes each channel through a single, channel-specific emission distribution, defined as $\psi_i(\mathbf{o}) = \mathcal{N}(\mathbf{o}; \boldsymbol{\mu}_i, \Sigma_i)$, with $\Sigma_i$ diagonal.
In particular, this benchmark can be seen as a degenerate case of the joint \gls{hmm} framework with a single state, i.e., $Z=1$.
To implement the benchmark, we first estimate the emission density $\psi_i^{\text{base}}(\mathbf{o})$ from the baseline windows $0, 1, \dots, w^* - 1$.
Hence, we estimate the emission density $\psi_{i,w}^{\text{ictal}}(\mathbf{o})$ associated with each window $w$ of the ictal phase, considering the windows $w^*, w^*+1, \dots, w$.
The abnormality score of the benchmark is then obtained by computing the \gls{kl} divergence between $\psi_i^{\text{base}}(\mathbf{o})$ and $\psi_{i,w}^{\text{ictal}}(\mathbf{o})$, and then averaging across the ictal windows $w = w^*, \dots, W-1$.



\section{Results}
\label{sec:result}

We consider a dataset with a total of $30$ patients that underwent \gls{seeg} monitoring at the IRCCS Institute of Neurological Science of Bologna, Italy.
For each patient, we select a single seizure and extract a $200$~s \gls{seeg} epoch including a $170$~s pre-onset baseline and the first $30$~s of the ictal phase, with the electrographic seizure onset marked by a clinician.
Our pre-processing pipeline is the same as presented in \cite{mason2026desynchronization}: each signal is segmented into overlapping windows of duration $\tau=1$~s and $\Delta \tau=0.25$~s shift, leading to an observation sequence $o_i[0], \dots, o_i[W-1]$ for each channel $i$, with $W=797$.
To compute the observations, we consider $B=7$ frequency bands, i.e., delta ($1$--$4$~Hz), theta ($4$--$8$~Hz), alpha ($8$--$14$~Hz), beta ($14$--$30$~Hz), low-gamma ($30$--$70$~Hz), high-gamma ($70$--$150$~Hz), and broadband ($0$--$250$~Hz). 


We consider as reference for \gls{ez} localization the \emph{thermocoagulated} channels, i.e., those channels whose recording sites underwent radiofrequency thermocoagulation at the end of the \gls{seeg} monitoring, and thus represent the initial surgical target identified by clinicians.
Hence, we compare our framework against the \emph{benchmark} approach presented at the end of Sec. \ref{sec:jointhmm}, in classifying epileptogenic vs non-epileptogenic channels. 
We measure classification performance using the \gls{auroc}, the \gls{auprc}, and the best F1, defined as the maximum F1-score over the evaluated decision thresholds.
These metrics are computed separately for each of the $30$ patients and summarized throughout the cohort through their mean and standard deviation.

To evaluate the framework robustness, we vary the strategy for estimating $A_{i,w}^{\text{ictal}}$ (sliding vs cumulative), the approach for computing the abnormality score (transition KL vs occupancy KL), the number of states $Z \in [1, 7]$ of the joint \gls{hmm} model, the number $L \in \{10, 20, 40\}$ of windows used for estimating $A_{i,w}^{\text{ictal}}$, and the feature set composition, which determines the size $F \in \{7, 14, 21\}$ of the observations $o_i[w] \in \mathbb{R}^{F}$.


\subsection{Illustrative example}
\label{subsec:ch_dynamics}

\begin{figure}[!t]
  \centering
\includegraphics[width=0.9\columnwidth]{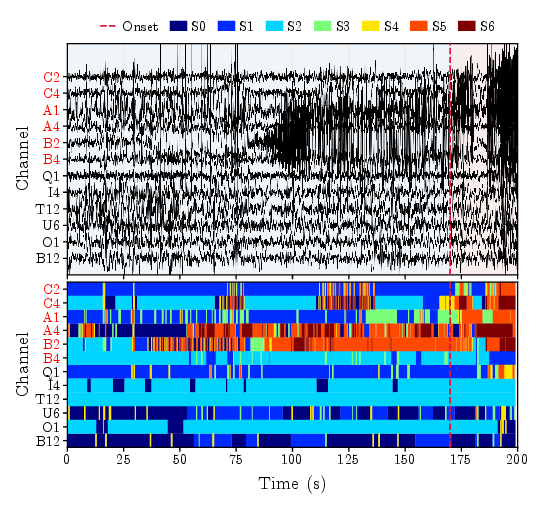}
  \caption{SEEG signal (top) and Viterbi-decoded state sequences (bottom) for a representative patient, with $Z=7$; channel labels in red belong to the EZ. \vspace{-0.4cm}
  }
  \label{fig:viterbi}
\end{figure}

We first consider a seizure from a representative patient, using only the energy-based features and setting the number of states to $Z=7$.
Fig.~\ref{fig:viterbi} shows the raw \gls{seeg} signals and the corresponding Viterbi-decoded state sequences for a subset of epileptogenic (i.e., part of the \gls{ez}) and non-epileptogenic channels, with the seizure onset starting at $t^*=170$ s.
Interestingly, our framework reveals that the \gls{ez} channels A4 and B2 already differ from the others during the baseline, visiting critical states that, for the other channels, appear only during the ictal phase.
After seizure onset, channels C2, C4, and A1 start visiting the same critical states more frequently, whereas non-epileptogenic channels such as T12 and I4 remain in normal states throughout the entire recording.
These patterns illustrate differences in channel dynamics and motivate the comparison of their transition probabilities.

\begin{figure}[!t]
  \centering
\includegraphics[width=0.9\columnwidth]{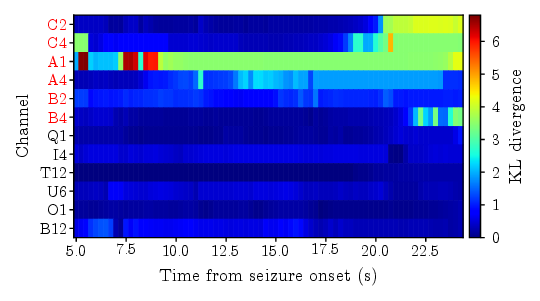}
  \caption{Transition KL divergence values, with $L=40$, and $Z=7$, for the
  same channels of Fig.~\ref{fig:viterbi}; channel labels in red belong to the EZ.\vspace{-0.3cm}
  }
  \label{fig:klheatmap}
\end{figure}

Fig.~\ref{fig:klheatmap} reports the abnormality scores $D_i(w)$ of each channel during the ictal phase, obtained with the transition KL and the sliding strategy ($L=40$ windows).
In particular, channel A1 exhibits the highest divergence immediately after seizure onset, while C2 and C4 show a marked increase approximately $20$ s later, followed by B4 towards the end of the interval, while all channels outside the \gls{ez} retain a low divergence throughout the seizure.
The timing and magnitude of the departure from the baseline dynamics thus vary across \gls{ez} channels.
A possible explanation is that some channels (including those on electrode B) already manifest epileptogenic abnormalities during the inter-ictal period.
In addition, since we consider only the first $25$ s of the ictal phase, some epileptogenic channels may be recruited into the seizure later.

\subsection{Performance Analysis}
\label{subsec:features}

\begin{figure}[!t]
  \centering
\includegraphics[width=\columnwidth]{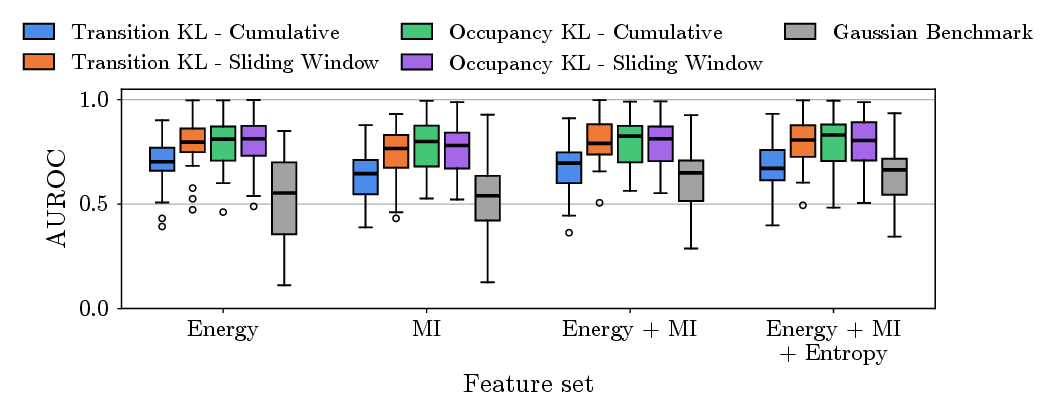}
  \caption{AUROC across patients for the different feature sets.\vspace{-0.3cm}}
  \label{fig:auroc_vs_feature_set}
\end{figure}

We now analyze the performance of our framework across the entire cohort of patients, comparing the different configurations of the proposed detection algorithm.
Fig.~\ref{fig:auroc_vs_feature_set} reports the performance distribution for different compositions of the feature set, with the number of features varying from $F=7$ (in the case of only energy or \gls{mi} being used) to $F=21$ (when all features are used).
The number of states is fixed to $Z=7$ while the estimation horizon is set to $L=20$. 

Interestingly, the \gls{hmm}-based variants are marginally affected by the choice of the feature set.
Using energy features only, \emph{occupancy KL - cumulative} and \emph{transition KL - sliding} achieve a mean \gls{auroc} of $0.790$ and $0.788$, respectively, and the inclusion of \gls{mi} does not lead to any significant changes. Conversely, the Gaussian benchmark strongly depends on the observation space, as its mean \gls{auroc} increases from $0.529$ with energy only, i.e., close to a random classifier, to $0.653$ when all the features are included.
Hence, the proposed joint \gls{hmm} framework can successfully classify epileptogenic channels by exploiting the temporal evolution of the energy features, while the benchmark needs additional features to compensate for the lack of temporal modeling.

We focus on the scenario where the feature set includes only the energy and the mutual information ($F=21$). 
As shown in Tab.~\ref{tab:results}, occupancy KL - cumulative achieves the highest mean \gls{auroc}, \gls{auprc}, and Best F1, i.e., $0.797$, $0.437$, and $0.511$, respectively, and is closely followed by transition KL - sliding and occupancy KL - sliding. 
Conversely, Transition KL - Cumulative does not significantly improve on the benchmark.
According to a paired two-tailed t-test across patients, considering a significance level of $0.05$ and uncorrected $p$-values, these three variants significantly outperform the benchmark in all performance metrics ($p < 0.01$), while no significant difference is observed among them.
In particular, occupancy KL - cumulative outperforms the benchmark in $27$ out of $30$ patients, whereas the benchmark performs worse than a random classifier (\gls{auroc}$=0.5$) in six patients.


\begin{table}[!h]
  \centering
  \caption{Mean $\pm$ std.\ deviation of AUROC, AUPRC and Best F1 across
  patients ($Z=7$).}
  \label{tab:results}
  \renewcommand{\arraystretch}{1.3}
  \resizebox{\columnwidth}{!}{%
  \begin{tabular}{lccc}
    \hline
    \textbf{Method} & \textbf{AUROC} & \textbf{AUPRC} & \textbf{Best F1} \\
    \hline
    Transition KL - Cumulative     & 0.688 $\pm$ 0.123 & 0.316 $\pm$ 0.168 & 0.404 $\pm$ 0.136 \\
    Transition KL - Sliding Window & 0.794 $\pm$ 0.117 & 0.415 $\pm$ 0.190 & 0.508 $\pm$ 0.139 \\
    Occupancy KL - Cumulative      & \textbf{0.797 $\pm$ 0.121} & \textbf{0.437 $\pm$ 0.197} & \textbf{0.511 $\pm$ 0.153} \\
    Occupancy KL - Sliding Window  & 0.790 $\pm$ 0.125 & 0.426 $\pm$ 0.197 & 0.497 $\pm$ 0.159 \\
    Gaussian Benchmark             & 0.635 $\pm$ 0.145 & 0.303 $\pm$ 0.146 & 0.404 $\pm$ 0.131 \\
    \hline
  \end{tabular}%
  }
\end{table}

Hence, the performance depends not only on the feature set used to characterize the latent dynamics, but also on the temporal strategy used to estimate them.
For transition KL, the sliding-window strategy raises the mean \gls{auroc} from $0.688$ to $0.794$ ($p < 0.001$), whereas for occupancy KL, the cumulative strategy performs slightly better ($0.797$ vs.\ $0.790$, $p = 0.40$).
Hence, detecting epileptogenic patterns while using transition probabilities leads to degraded results when the \gls{hmm} model is estimated over the whole ictal interval, whereas using stationary state distribution, which summarizes the time spent in each state, is less sensitive to the interval used to estimate the latent model's parameters.


\begin{figure}[!t]
  \centering
\includegraphics[width=\columnwidth]{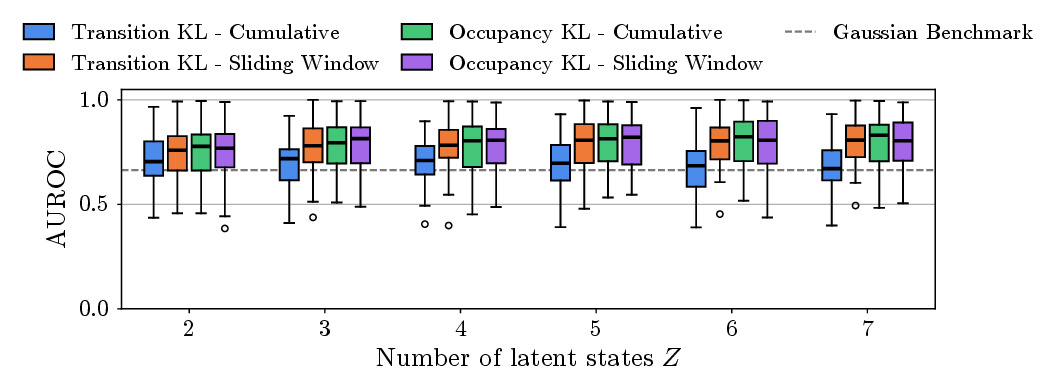}
  \caption{AUROC across patients versus number of states $Z$.\vspace{-0.3cm}}
  \label{fig:auroc_vs_K}
\end{figure}

\begin{table}[!h]
  \centering
  \caption{Mean AUROC across patients of the sliding-window variants as a
  function of $L$ and $Z$.}
  \label{tab:auroc_vs_L}
  \renewcommand{\arraystretch}{1.3}
  \resizebox{\columnwidth}{!}{%
  \begin{tabular}{ccccccc}
    \hline
     & \multicolumn{3}{c}{\textbf{Transition KL - sliding}} & \multicolumn{3}{c}{\textbf{Occupancy KL - sliding}} \\
    \cline{2-4} \cline{5-7}
    $\boldsymbol{Z}$ & $\boldsymbol{L=10}$ & $\boldsymbol{L=20}$ & $\boldsymbol{L=40}$ & $\boldsymbol{L=10}$ & $\boldsymbol{L=20}$ & $\boldsymbol{L=40}$ \\
    \hline
    2 & 0.730 & 0.737 & \textbf{0.740} & 0.736 & 0.740 & \textbf{0.742} \\
    3 & 0.765 & \textbf{0.774} & 0.766 & \textbf{0.776} & 0.775 & 0.770 \\
    4 & 0.767 & \textbf{0.782} & 0.779 & 0.776 & \textbf{0.777} & 0.771 \\
    5 & 0.777 & \textbf{0.786} & \textbf{0.786} & 0.785 & \textbf{0.788} & 0.780 \\
    6 & 0.772 & \textbf{0.787} & 0.786 & 0.782 & \textbf{0.785} & 0.779 \\
    7 & 0.777 & \textbf{0.794} & 0.793 & 0.784 & \textbf{0.790} & 0.785 \\
    \hline
  \end{tabular}%
  }
\end{table}

We finally investigate whether our results depend on the choice of the hyperparameters.
Fig.~\ref{fig:auroc_vs_K} shows the distribution of \gls{auroc} across patients as a function of the number $Z$ of states.
Both the sliding-window approaches and occupancy KL- cumulative outperform the benchmark for all values of $Z$, with a median \gls{auroc} between $0.77$ and $0.80$ for $Z \geq 3$; instead, the KL cumulative transition tends to decrease as $Z$ increases, consistently with the attenuation effect discussed above.
Similarly, Tab.~\ref{tab:auroc_vs_L} analyzes whether the sliding-window approaches are affected by the choice of $L$.
These results show variations of the mean \gls{auroc} below $0.02$ for every $Z$, with $L=20$ achieving the highest mean \gls{auroc} in most cases.

\section{Conclusion}
\label{sec:conc}

In this work, we proposed a new framework that models \gls{seeg} signals as a joint \gls{hmm}, where each channel is associated with its own transition probability matrix, while all channels share the same state and observation spaces.
The proposed methodology allows us to decode the latent state sequence associated with each channel and time point, potentially revealing the transition dynamics associated with epileptiform activity and seizure generation.

We defined an abnormality score for each channel based on the divergence between its temporal dynamics during the baseline and at the seizure onset.
We then compared our methodology against a benchmark approach that does not consider state transitions to classify epileptogenic channels. 
Our results indicate that capturing latent state evolution, as done by our model, improves the characterization of the \gls{ez}.
The proposed framework has hence the potential to support the clinical interpretation of \gls{seeg} signals and to improve our understanding of the hidden dynamics underlying seizure generation in drug-resistant epilepsy.

\bibliographystyle{IEEEtran}
\bibliography{biblio.bib}

\end{document}

%% file: acronyms.tex
\newacronym{iot}{IoT}{Internet of Things}
\newacronym{ml}{ML}{machine learning}
\newacronym{ez}{EZ}{Epileptogenic Zone}
\newacronym{seeg}{SEEG}{Stereo-Electroencephalography}
\newacronym{lvfo}{LVFOs}{Low-Voltage Fast Oscillations}
\newacronym{stft}{STFT}{Short Time Fourier Transform}
\newacronym{dft}{DFT}{Discrete Fourier Transform}
\newacronym{wt}{WT}{Wavelet Transform}
\newacronym{ei}{EI}{Epileptogenic Index}
\newacronym{hmm}{HMM}{Hidden Markov Model}
\newacronym{meg}{MEG}{Magneto-Electroencephalography}
\newacronym{kl}{KL}{Kullback–Leibler}
\newacronym{mi}{MI}{Mutual Information}
\newacronym{em}{EM}{Expectation-Maximization}
\newacronym{bw}{BW}{Baum–Welch}
\newacronym{auroc}{AUROC}{Area Under the Receiver Operating Characteristic Curve}
\newacronym{auprc}{AUPRC}{Area Under the Precision-Recall Curve}